# Multiplex Networks Provide Structural Pathways for Social Contagion in Rural Social Networks

Yongren Shi[1*], Edo Airoldi[2], and Nicholas A. Christakis[3,4,5]

[1] Department of Sociology, University of Arizona, Tucson, AZ 85721, USA.

[2] Department of Statistical Science, Fox School of Business, Temple University, Philadelphia, PA 19010, USA

[3] Yale Institute for Network Science, Yale University, New Haven, CT 06520, USA.

[4] Department of Sociology, Yale University, New Haven, CT 06520, USA.

[5] Department of Statistics and Data Science, Yale University, New Haven, CT 06520, USA.

*To whom correspondence should be addressed. e-mail: yongren@arizona.edu

**Competing Interest Statement:** The authors declare that they have no competing interests.

**Keywords:** Tie Multiplexity; Social Networks; Health Intervention; Social Contagion; Network Torque.

**Acknowledgments**
We thank our field team in Honduras for collecting the data; the Honduras Ministry of Health for providing advice regarding, and approval of, intervention materials and study instruments; and the Inter-America Development Bank, World Vision Honduras, Child Fund Honduras, and Dimagi, for delivering the intervention. Rennie Negron supervised the overall execution of the field data collection and Liza Nicoll managed the data. Mark McKnight and Wyatt Israel developed the Trellis software to map networks and collect the data. Tom Keegan provided overall project management, and Maria Vassimon de Assis, Will Oles, Petergaye Murray, and Eric Liu provided research assistance. The first author also thanks Freda Lynn for helpful

**Abstract**

Human social networks are inherently multiplex, comprising overlapping layers of relationships. Different layers may have distinct structural properties and interpersonal dynamics, but also may interact to form complex interdependent pathways for social contagion. This poses a fundamental problem in understanding behavioral diffusion and in devising effective network-based interventions. Here, we introduce a new conceptualization of how much each network layer contributes to critical contagion pathways and quantify it using a novel metric, "network torque." We exploit data regarding sociocentric maps of 110 rural Honduran communities using a battery of 11 name generators and an experiment involving an exogenous intervention. Using a novel statistical framework, we assess the extent to which specific network layers alter global connectivity and support the spread of three experimentally introduced health practices. The results show that specific relationship types—such as close friendships—particularly enable non-overlapping diffusion pathways, amplifying behavioral change at the village level. For instance, non-redundant pathways enabled by closest friends can increase the adoption of correct knowledge about feeding newborns inappropriate "chupones" and enhance attitudes regarding fathers' involvement in postpartum care. Non-overlapping multiplex social ties are relevant to social contagion and social coherence in traditionally organized social systems.

**Keywords:** Tie Multiplexity; Social Networks; Health Intervention; Social Contagion; Network Torque.

**Introduction**

Traditionally, human social networks are localized (Watts 1999, 2004; Apicella et al. 2012), especially in rural settings, with different clusters of familial and friendship ties even within the same community (Entwisle et al. 2007). Yet, we have an incomplete understanding of how different types of relationships might relate to each other or how any overlap in types of ties (known as "multiplexity") might alter the structure and function of the local network (Verbrugge 1979; Smith and Papachristos 2016; Shakya, Christakis, and Fowler 2017). For instance, a tie to a close friend might overlap with a tie to the same person from whom one might borrow money, but a tie to someone with whom one spends free time might not overlap with a tie to someone from whom one seeks health advice. This may matter for processes undergirding social contagion, social coherence, and social capital creation.

At the dyadic level, distinct expectations are associated with different types of relationships, which in turn can affect attributes of the tie, such as its intimacy, frequency, and duration (Fischer 1977; Verbrugge 1979). These qualities of the multiplex ties can also influence the rate of contagion across them, e.g., the extent to which a piece of information can pass between a pair of connected individuals. For instance, gossip tends to spread within dyads connected by a strong and multiplex tie (Centola and Macy 2007), but the diffusion of unpopular norms often relies on weak ties (Cowan 2014). At the global network level, when multiple types of relationships work together, they can create new and otherwise unlikely pathways for social contagion. Overlooking this tie multiplexity in network mapping can obscure the importance of each layer and limit the effectiveness of pathway-based interventions.

Moreover, interventions seeking to leverage social contagion often involve the deliberate exploitation of the structure and function of social networks in order to generate social influence and accelerate population-wide behavior change. Identifying and targeting structurally influential individuals for information dissemination is an increasingly recognized and effective strategy, and it has been used in many contexts (Bond et al. 2012; Centola 2010; Chami et al. 2017; Kim et al. 2015; Muchnik, Aral, and Taylor 2013; Paluck, Shepherd, and Aronow 2016; Rand, Arbesman, and Christakis 2011; Valente 2012; Valente et al. 2003; Valente and Davis 1999), including in rural regions of developing countries (Kim et al. 2015; Airoldi and Christakis 2024; Shridhar, Alexander, and Christakis 2021; Banerjee et al. 2013; Alexander et al. 2022; Kelly et al. 2006). However, effectively generating large-scale cascades in population-wide social networks requires the existence of ample network pathways among people in a population by which information or behavior can spread to distant sections of the network.

Here, we seek to understand and estimate the impact of multiplex networks on contagion – whether naturally occurring or experimentally induced. To do so, we explore experimentally induced contagions of health knowledge, attitudes, and practices in 110 isolated villages in the Department of Copán, Honduras. To collect multiplex social network data, we used bespoke (publicly available) software (Lungeanu et al. 2021) and a battery of "name generator" questions to capture three categories of social relationships among individuals within each village (see **Table 1** for a list). The first category includes role relations, such as parent and child, spouse, sibling, and "patron." The second category is social activities, including choosing someone to confide in regarding personal and private matters, to spend free time with, or as closest friends. The third category involves exchange relations, including trust to lend/borrow money and seeking and giving health advice. Here, we use measures that were collected at baseline (prior to the 22-month-long maternal and child health intervention) in roughly 2016 to construct the multiplex networks.

[Table 1 is here]

We focus on two objectives. First, to what extent does a network layer defined by one type of relationship add additional pathways in a village network compared with other layers? We propose a novel metric, *network torque*, that measures the degree to which a network layer alters the underlying pathways between pairs of nodes. The measure has also direct empirical relevance for devising network interventions that target specific types of relationships in order to facilitate contagion. Second, how does the formation of pathways enabled by a particular network layer contribute to social contagion? The difficulty of this question arises from the intertwining of multiple network layers, which collectively shape pathways. We devise a method that uses tie directionality to evaluate the influence of network layers on experimentally induced social contagions. The results indicate substantial variation in the importance of different network layers in facilitating social contagion. In particular, the "closest friend" layer plays a significant role in facilitating village-wide knowledge diffusion through the crosscutting pathways it enables.

**Network Multiplexity Structuring Social Contagion**

Different expectations arise from the social context and relational category in which the tie is situated—such as kinship (Fischer 1977), friendship (Fischer 1982; Verbrugge 1979), co-offending association (Papachristos 2009; Smith and Papachristos 2016; Hsiao, Leverso, and Papachristos 2023), or acquaintance (Becker et al. 2020)—and they influence how individuals interact and the content of communication. For instance, kin-based ties often carry normative obligations of mutual aid, sustained emotional involvement, and frequent contact (Fischer 1977), while workplace ties, for example, are shaped by task-oriented exchanges, role differentiation, and hierarchical constraints (Blau 1964; Podolny and Baron 1997). Similarly, friendships are typically marked by voluntary engagement, reciprocity, and socioemotional support (Fischer 1982), whereas acquaintanceships may be characterized by limited intimacy but serve as conduits for novel information (Granovetter 1973; Small 2013).

These relational expectations directly influence dyadic attributes such as intimacy, trustworthiness, legitimacy, and temporal stability, which in turn affect interpersonal dynamics across the network. High-intimacy, frequent-contact ties may foster rapid emotional exchange, mutual reinforcement of attitudes, and greater susceptibility to influence, while low-intimacy, infrequent-contact ties may encourage selective disclosure and limit emotional contagion (Brashears 2011; Marsden and Campbell 1984). The implications for diffusion processes are substantial: different types of relationship can produce markedly different contagion patterns. For example, gossip—a form of socially sensitive information—tends to propagate most effectively within dyads characterized by strong, enduring, and socially multiplex connections, where trust and repeated interaction increase the likelihood of transmission (Centola and Macy 2007; Centola 2018). By contrast, the spread of behaviors or norms that may be socially costly or unpopular often relies on weaker or less emotionally embedded ties, which act as bridges between otherwise disconnected network clusters and reduce reputational risk for adopters (Cowan 2014; Kitts 2003).

Network multiplexity not only affects dyadic interpersonal dynamics, but also directly influences social diffusion by shaping the pathways for contagion on population-wide networks. Consider a stylized example of pathways (**Figure 1**) that are constructed by three types of relationships: family (network layer $\alpha$); people with whom one reports being able to discuss personal and private matters (network layer $\beta$); and people with whom one shares health advice (network layer $\gamma$). Given that a piece of health information, for example, enters the system at individual *s*, its diffusion depends on the interdependence of the three layers. On the one hand, layers $\alpha$ and $\beta$ have a large degree of tie overlap, forming redundant channels for information to spread (incidentally, all the social ties in layer $\alpha$ are subsumed in layer $\beta$, so $\beta$ renders $\alpha$ redundant, but the opposite does not apply). Layer $\gamma$, on the other hand, bridges otherwise separate parts of the community, and it becomes a critical constituent of the pathway on which the information can spread between *s* and *t*.

[Figure 1 is here]

As shown in the illustration, the synergistic interaction of multiplex ties generates non-reducible contagion pathways that can alter diffusion dynamics. These pathways emerge collectively from the interaction of ties across distinct relational layers, creating connectivity that cannot be reduced to any single-layer topology (Boccaletti et al. 2014). The superimposition of multiple networks enables nodes or edges with multiplex connections to develop higher-order structural roles, facilitating shortcuts across different types of networks (Kivelä et al. 2014).

Multiplex networks also demonstrate topology-function (mis)alignment, where specific relational layers disproportionately enable the diffusion of particular practices based on their functional capacity. Multiple mechanisms may be responsible for this alignment. For instance, certain types of relationships are institutionalized in ways that are conducive to spreading specific information or behaviors through networks. Health practices requiring normative validation (e.g., vaccination) propagate most effectively through layers with institutional authority (e.g., professional networks), where relationships to health authorities provide both connectivity and legitimacy (Small 2017; Van den Bulte and Lilien 2001). Conversely, practices demanding affective commitment (e.g., dietary changes) diffuse through close/kinship ties where strong ties enable normative reinforcement (Christakis and Fowler 2007; Perry and Pescosolido 2010, 2015). Thus, the structural positions of these functional ties—whether being embedded in redundant local clusters or serving as bridges between distinct network clusters—can influence both the speed and the reach of diffusion.

This alignment challenges monoplex assumptions that centrality or tie strength universally predicts influence. A node's diffusion efficacy depends on whether its structural position aligns with the practice's functional requirements: a central professional contact may drive technological adoption but lack influence over smoking cessation, which may flow through friendship layers (Rawlings et al. 2023). Effective interventions must therefore map the specific topology-function coupling for each practice rather than targeting generic "influencers." Ignoring multiplexity risks designing interventions based on an incomplete map, targeting inefficient pathways or overlooking crucial leverage points where the interaction of relational layers creates the most potent conduits for change (Valente 2012).

While there are extensive research on the structural and dynamic properties of multiplex networks (Bianconi 2018; Boccaletti et al. 2014; Kivelä et al. 2014), demonstrating the influence of overlapping networks on the generation or disruption of system behavior, including resilience to random failures (De Domenico et al. 2014; Liu, Stanley, and Gao 2016), percolation (Parshani, Buldyrev, and Havlin 2010; Zhao and Bianconi 2013), cascades (Brummitt, D'Souza, and Leicht 2012; Buldyrev et al. 2010), and network control (Menichetti, Dall'Asta, and

Bianconi 2016), existing studies have two main limitations. First, they primarily analyze natural or physical networked systems using synthetic data or simulations, rather than real face-to-face social networks. Second, they lack the development of estimation strategies to assess the impact of network layers on social contagion within the networks, with exogenous introduction of information in particular. This study intends to address these limitations.

**The Social Network Intervention in Rural Honduras**

To investigate and estimate how the topology and function of network layers interact to facilitate social contagion, we leverage an intervention study promoting prenatal and postnatal health care practices and knowledge in rural Honduras. Despite progress made in reducing neonatal, infant, and child mortality, Honduras continues to face challenges with one of the highest neonatal mortality rates in Latin America. In 2008, neonatal deaths accounted for 51% of all deaths in children under 5 years old, with 40% attributed to premature labor and another 40% to asphyxia and infection. Although 79% of neonates initiate breastfeeding within an hour of birth, only 30% are exclusively breastfed for the recommended first 6 months of life (Shakya, Stafford, et al. 2017; UNICEF Country profile 2012) with many instead being fed a local neonatal foodstuff known as "chupones."

We conducted this study in the Department of Copán, Honduras, in an area of over 200 square miles of rugged mountainous terrain. Copán is a rural region in which geographic barriers and limited transportation infrastructure tend to isolate villages from one another. People in this area live on less than about two dollars a day. See **SI Appendix A.**

The intervention study consists of multiple stages. We first conducted a photographic census in 176 villages in 2015. Recruitment rates were high: government census data show approximately 32,500 eligible individuals in these villages, of whom at least 94.8% (N=30,815) agreed to enroll in our own *de novo* census (in which we photographed all village residents and collected basic data about them). Of these people, 24,702 people (in a total of 10,013 households) then enrolled in our longitudinal study. They agreed to participate in a baseline survey (in 2016), undergo randomization, receive a 22-month multi-faceted maternal and child health intervention (if selected as a target household), and be surveyed approximately annually or biannually. The survey includes a battery of "name generator" questions to capture social relationships (Marsden 1990). The terrain and insular villages of Copán ensured that the vast majority of face-to-face social ties existed within, rather than between, villages. We developed a specialized (publicly available) mobile application "Trellis," specifically designed for conducting a photographic network census, to gather survey data (Lungeanu et al. 2021). By capturing individual photographs, this app enables participants to visually identify their social contacts, thereby improving the accuracy of network data collection. The village populations (adolescents and adults) ranged from 41 to 509 individuals, and the average household size was 2.5 (±1.4) (See **SI Appendix B** for village and network descriptive statistics). At baseline, the average age of the 24,702 participants was 33 (SD=17; range: 11-93); 58% were women; and 59% were married.

The parent randomized controlled trial (RCT) (Airoldi and Christakis 2024; Shakya, Stafford, et al. 2017), we evaluated (1) what fraction of households in a community needs to be targeted in order to maximally change knowledge, attitudes, and practices with respect to various health outcomes; and (2) whether a special network targeting algorithm is more effective than a control strategy of random targeting. We randomized the 176 villages using an 8×2 factorial design in which we varied the *percentage* of people targeted per village (0%, 5%, 10%, 20%,

30%, 50%, 75%, and 100%), and (independently) the method of choosing the target households (*random* targeting and *friendship nomination* targeting). At the extremes, no households (0%, in 22 villages) or all households (100%, in 22 villages) were selected for intervention; otherwise, there were 11 villages in each cell (see **Table S1**). The present study uses data from 110 villages that received random targeting and villages that received 0% and 100% interventions. The remaining 66 villages that received a special network targeting algorithm were excluded here.

In the 110 villages, a health intervention program was implemented in a total of 2,439 random households, with 6,019 individuals. Each targeted household had the potential to receive multiple intervention counseling sessions over the course of 22 months (November 2016 – August 2018). It was comprised of a series of household-level counseling sessions targeting maternal, child and neonatal practices using 15 distinct educational modules. It was delivered by trained community health workers on a monthly basis to the households selected for the study.

The educational package used the Timed and Targeted Counselling (ttC) methodology complemented with alternative methods of face-to-face communication including songs, rhymes and riddles, designed with World Vision and Child Fund Honduras (International World Vision 2015; Shakya, Stafford, et al. 2017). The social and behavior change communication strategy for the intervention was designed using the "P-Process," a tool developed by Johns Hopkins Center for Communication Programs (Collaborative HCC 2016). This methodology uses narrative and negotiation in a 1-2-hour visit with families to discuss positive and negative scenarios and create a list of agreements with families to try out new practices. Counselling is offered to all family members, tailored to the specific stage of pregnancy or age of the child. Counseling was extended to all members of the selected households, regardless of whether they had young children or pregnant women residing within. The educational package was implemented to a total of 2,439 random households and 6,019 people in the 110 villages.

To provide an example, in the statistical analysis of social contagion pertaining to the knowledge item regarding the use of a chupon (pacifier) for infants under 6 months old, this particular item was covered within educational modules specifically addressing breastfeeding. In our analysis, individuals who received counseling sessions specifically covering breastfeeding-related topics would be classified as having received the intervention. However, individuals who received counseling sessions focused on other topics unrelated to breastfeeding would not be considered as having received the intervention for this particular knowledge item.

A follow-up survey on attitude, knowledge and practices ("Wave 3") was conducted in 2019 when all intervention activities had concluded, about 24 months after the delivery of the intervention first began.

**Structural Variation of Village Multiplex Networks**

Our study area spans over 200 square miles characterized by rugged mountainous terrain. The presence of geographic barriers and limited transportation infrastructure contribute to the isolation of villages within the region. We sociocentrically mapped the networks within each village, aiming to identify relevant structural properties of different relationships.

The types of relationships show significant structural variation. **Figure 2** displays network visualizations of six relationship types from a selected village, rendered in isolation from other relationships (upper row) and superimposed onto the composite network consisting of all 11 types of relationships, which is represented by the color gray (lower row). Significant differences are evident by visual inspection alone. For instance, the network layer composed of closest friends exhibits a structure that is locally cohesive, e.g., "a friend of a friend is also my friend" (Isakov et al. 2019), but globally expansive, e.g., distant nodes can be connected through a path. This indicates that a piece of information might easily navigate through the pathways formed by these relationships alone. In contrast, the networks focusing on getting health advice and borrowing money display multiple but weakly connected long chains of interactions, implying that exchanges primarily depend on connections with trustworthy individuals locally, not with distant others with whom network shortcuts and closed loops may form. In addition, as expected, the sibling network is characterized by a high level of local clusters with few connections in-between.

[Figure 2 is here]

Network metrics of village networks confirm the observations. **Figure 3** presents the distributions of network characteristics, including prevalence (proportion of ties), clustering coefficient (Watts and Strogatz 1998), and reachability (Moody and White 2003), across village networks. For example, the "closest friend" layer has moderate local clustering across villages, but highest reachability, i.e., the degree to which two individuals can be mutually reached within a network. On average, 76.7 percent of the individuals can be connected through a pathway composed exclusively of closest friends. Networks composed of "free time" and "personal private" also have relatively high levels of reachability. In contrast, the "health advice give" network shows local clustering similar to that of closest friends, yet its reachability is much reduced, at 27.9%, pointing to a potent echo chamber of health information within local clusters and difficulties in its ability to spread information to distant network sections.

[Figure 3 is here]

There are also visible distinctions in the distributions of relationships in the composite networks (the lower row in **Figure 2**). For instance, on the one hand, relationships involving spending free time together, health advice, trusting to borrow money, and sibship demonstrate a high level of overlap within the local clusters but with few connections between clusters. On the other hand, the pattern of "closest friend" layer confirms that it not only reinforces existing local connections, but also tends to span across distant sections of the social network. In summary, different relationships exhibit varying capacities to restructure the underlying overall social network, which may result in various levels of non-redundant pathways.

**Network Torque and its Structural Correlates**

We can formalize the foregoing intuition in a new network metric, *network torque*. Network torque is conceptualized as the extent to which a particular network layer contributes to the formation of short pathways connecting distant nodes in a village (or any otherwise bounded population), e.g., $s$ and $t$ in **Figure 1**. Put differently, we quantify the impact of a network layer by assessing its ability to disrupt the shortest pathways within the network if that particular type of relationship were to be removed. We define a network layer as *critical* to a pair of $k$-degree nodes (i.e., connected by a shortest path of $k$ consecutive edges) in the composite network when its absence would require the two nodes to be traversed through a longer route or become completely disconnected; and similarly, a network layer is *non-critical* to a shortest pathway when it can be removed without affecting the current length of paths connecting the two nodes.

**Figure 4a** illustrates critical and non-critical pathways. The criticality of a layer $L$ to a pair of $k$-degree nodes $i$ and $j$ can be expressed as following:

$$\sigma_{i,j}(L) = \begin{cases} 1, & d_{i,j}(E \neg L) > d_{i,j}(E) \\ 0, & d_{i,j}(E \neg L) = d_{i,j}(E) \end{cases}$$

where $d_{i,j}(E)$ represents the shortest distance between nodes $i$ and $j$ in the composite social network, $E$, and $d_{i,j}(E \neg L)$ represents the shortest distance in the network with layer $L$ being removed. For example, in **Figure 1**, the family layer $\alpha$ is non-critical to the pair of 5-degree separated nodes $s$ and $t$ as the pathway is resilient to the removal of layer $\alpha$; and $d_{s,t}(E) = 5$. If layer $\alpha$ is removed, the length of the shortest path between $s$ and $t$ remains 5; and $d_{s,t}(E \neg \alpha) = 5$. Therefore, the criticality of layer $\alpha$ to $s$-$t$ is 0. In contrast, the same pair of nodes, $s$ and $t$, is not resilient to the removal of relationships dedicated to discussing personal and private matters (layer $\beta$) or sharing health information (layer $\gamma$), so $\sigma_{s,t}(\beta) = 1$ and $\sigma_{s,t}(\gamma) = 1$. We consider the measure of network torque of layer $L$ as the proportion of all node pairs in a network that are not resilient to the removal of the layer $L$:

$$t_L = \sum_{i \neq j} \sigma_{i,j}(L) / \sum_{i \neq j} \sigma_{i,j}$$

where $\sigma_{i,j}$ is a binary indicator of whether nodes $i$ and $j$ are connected by a finite-length path in the composite network. Consider the impact of removing layer $\alpha$ on the pathway connecting $s$ and $t$ in **Figure 1**. Because edges in layer $\alpha$ are a subset of edges in layer $\beta$, the removal of layer $\alpha$ would not alter the distance between any pair of nodes $i$ and $j$, which means that the criticality of $\alpha$ to any node pair is zero; $\sigma_{i,j}(\alpha) = 0$. Therefore, $t_\alpha = 0$, indicating that layer $\alpha$ has no influence on the formation of non-redundant pathways. In contrast, layer $\beta$ and $\gamma$ have significantly greater impacts, with $t_\beta = 0.39$ and $t_\gamma = 0.56$, indicating that 39% and 56% of the node pairs would be connected by longer paths or disconnected if the respective layers are removed. This measure ranges from 0 to 1, with a high value presumed to suggest a great likelihood for information, say, to spread through the available pathways created by a given type of relationship.

[Figure 4 is here]

**Figure 4b** shows that different relationship types exhibit varying levels of network torque. Social relationships of "partner," "parent" and "patron" have a negligible impact on the creation of crosscutting connections, as expected. In contrast, the "closest friend" relationship exerts the highest torque on networks with an average of 0.349, indicating that its removal would disconnect at least 34.9% of the shortest pathways. And interestingly, its impact in creating new pathways is 3.3 and 9.4 times higher than the other two friendship name generators, "free time" and "personal private." Another notable relationship is siblings, which demonstrates a higher network torque compared to all other relationships, except for the "closest friend." Despite sibling networks exhibiting a high degree of cliqueness (**Figure 2** and **3**), adult siblings are likely to have established their own families and friendships outside of their immediate family circle, serving as bridges between clusters.

Three structural characteristics regarding the distribution of relationships may collectively influence network torque. The first is the prevalence of a type of relationship in the network. When a type of relationship, e.g., with whom to spend free time, is common, they are likely to be present in the shortest pathways just by chance. **Figure 4c** illustrates a positive linear relationship between the prevalence of relationships and network torque (log-transformed).

However, the prevalence of a specific relationship does not necessarily indicate that its removal would have a sizable impact on the disruption of pathways. The persons with whom one spends free time may also be their parents and trusted individuals from whom they borrow money, therefore, removing the "free time" relationships is not likely to disconnect a large portion of the pathways on which it is involved.

The second structural characteristic is the degree to which a relationship type does not coexist with other types of relationships on a dyad (i.e., monoplexity). For example, monoplex ties accounts for 54.1% of closest friends. This is partly due to the design of the name generator, which explicitly precludes nominations of parents, partners, and siblings from being "closest friends." But it is also likely that individuals deliberately chose close friends with whom they do not share multiplex relationships. In contrast, social ties associated with spending free time tend to exist within multiplex relationships. **Figure 4d** shows that network torque is positively associated with the proportion of monoplex ties of each relationship type. Notably, the "patron" relationship is an exception. Patrons are individuals in villages who provide daily jobs to the respondents; 56% of patron nominations are monoplex relationships (ranking 3[rd]), but they have the lowest prevalence, accounting for only 0.6% of all nominations (ranking 11[th]).

We also expect a relationship layer's torque to be associated with its structural distribution. Specifically, high torque relationships are expected to be more likely to exist as bridges between nodes that are not connected through other shorter paths. Thus, the third measure we consider is edge betweenness centrality, which measures the number of shortest paths in a network that pass through an edge (we do not distinguish whether a said edge is multiplex or monoplex when calculating betweenness centrality) (Granovetter 1973). Edges with high betweenness centrality play a vital role in facilitating communication and information flow between distant sections of a network. **Figure 4e** confirms our expectation of a positive linear relationship between average edge betweenness centrality across all villages and network torque.

The three characteristics—prevalence, dyadic monoplexity, and bridgeness—are conceptually distinct and operate under independent logics of tie formation, but they collectively contribute to the likelihood of a relationship type establishing non-redundant pathways that serve as important channels for information transmission.

**Evaluating the Effect of Multiplex Networks in Spreading Health Knowledge Due to Exogenous Interventions**

Let us consider the relative importance of each network layer to social contagion. We can take advantage of a health intervention that was implemented in a random set of 6,256 households and 15,398 people residing in the 110 villages that were part of a larger RCT (Airoldi and Christakis 2024). Demographic and social network information and baseline health knowledge were collected during wave 1 (2016). The treatment contained monthly in-home counseling and education sessions for 22 months focusing on various target outcomes such as prenatal and maternal health, as well as prevention and management of diarrhea and respiratory illnesses. A follow-up survey (Wave 3) was conducted in 2019 when all intervention activities had concluded, about 24 months after the delivery of the intervention first began (there was also a Wave 2, not used here).

We can evaluate the extent to which, after an exogenous intervention was applied to random people within the village networks, a village-wide change in health knowledge is facilitated by the network pathways enabled by a particular type of relationships. Central to our evaluation is a model that predicts knowledge accuracy at Wave 3 based on the social influence calculated from all the 2-degree, 3-degree and 4-degree intervened alters who are connected to the focal individuals through *critical pathways* (i.e., only available due to the presence of a particular network layer). In other words, the model uses the social influence that passes through the critical pathways between the intervened individuals and the focal individuals.

Consider the illustration in **Figure 4a**. The *total* social influence received by the focal individual can be calculated as the number of unique *k*-degree alters who have received educational interventions at Wave 1, which is 4 in Figure 4a. We are interested in *a subset* of the influence that goes through the critical pathways enabled by a particular network layer ("blue" relationships), which are pathways A and E in the example. Path B is not a critical pathway as it is intact if the blue link is removed. Path C is not a critical pathway either as the removal of "blue" relationships wouldn't disconnect the alter and focal individual because an alternative pathway (D) is available. We count the numbers of intervened alters that are connected through *k*-degree critical pathways, and use them as variables for social influence.

A methodological approach facilitating our assessment is a framework for *counterfactual predictions*, comparing individual-level knowledge accuracy when social influence variables in the statistical model are set to their actual values versus when they are set to zero, corresponding to the removal of network layers in counterfactual scenarios. We calculate the *reduction* in the predictions of individuals' correct health knowledge between these two conditions.

Our estimation strategy is as follows: for each of the selected knowledge interventions, we first specify a logit model predicting knowledge accuracy at wave 3 based on the critical pathways enabled by a network layer:

$$logit(P(K_i^{w3} = 1)) = \beta_0 + \beta_1 K_i^{w1} + \beta_2 \xi_{i,L}^k + \beta_3 \xi_{i,-L}^k + \beta_4 N_i^k + \boldsymbol{X_i \delta} + \alpha_V + \epsilon_i$$

where $K_i^{w3}$ and $K_i^{w1}$ are binary indicators denoting individual *i's accuracy* on the selected knowledge items *K* during Wave 3 and Wave 1, respectively. Subscript *i* denotes individuals, and *L* denotes network layer. Coefficients $\beta_2$ and $\beta_3$ represent the "social contagion effects," wherein $\xi_{i,L}^k$ and $\xi_{i,-L}^k$ represent the numbers of *k*-degree intervened alters who are connected with the focal individuals through *critical pathways* enabled by the network layer *L*, and through the rest of the pathways $(-L)$. Including $\xi_{i,-L}^k$ prevents the omission of any *k*-degree pathways through other network layers that may influence the association of outcome and independent variable of interest, $\xi_{i,L}^k$. $N_i^k$ captures the total size of *k*-degree alters, regardless of their treatment status.

After the model is estimated, we evaluate the difference in the predicted outcomes when a particular network layer is included and when it is removed. By inserting the observed values of $\xi_{i,L}^k$, the predicted knowledge accuracy at Wave 3 was obtained as if the presence of a particular network layer were taken into consideration. Conversely, by setting $\xi_{i,L}^k$ as zeros, the predicted knowledge accuracy was obtained as if the particular network layer were *removed* in the counterfactual network. We compute the difference in the prediction probability for each of the individuals, and designate the average as the reduction in adoption when a network layer is removed.

Covariates $X_i$ include individual *i's treatment status*, *sociability* (i.e., the total number of network nominations, that is, the total number of ties, of whatever type, to all connected alters), *age*, *gender*, *education* (i.e., number of school years completed), *income*, and *self-reported health*. The control variable *income* is converted into a continuous variable ranging from 1 to 4, from a four-category variable, including: "not sufficient with major difficulties," "not sufficient with some difficulties," "sufficient," and "enough to live on and save." The model includes village fixed effects ($\alpha_V$); thus, the result is not biased by omitted village-level characteristics.

Here, we highlight three methodological considerations. First, we do not intend to causally estimate the network effect of a specific type of relationships net of other relationship types. This estimation task is challenging, if not impossible, as network pathways are typically shaped by multiple relationship types, with long chains of connections rarely formed by a single type. Instead, our strategy is to evaluate the "disruption" of network pathways in reducing social influence if a specific network layer is removed, where our conceptualization of critical pathways is a key component. However, our models do account for non-critical pathways in aggregate terms to avoid omitting alternative information flows that could bias the results.

Second, due to the absence of data tracing exact information transmission pathways, we evaluate the impact of network-layer removal using two sets of critical pathways. Primary critical pathways are restricted to intermediary individuals on the paths with accurate knowledge at Wave 3 or direct intervention exposure at Wave 1, acting as validated channels where information degradation is minimized. These pathways, though fewer, are theoretically more reliable for knowledge transmission—especially over long chains. Secondary critical pathways include all possible routes, even those involving individuals with incorrect knowledge, which may introduce noise.

We emphasize that our goal is not to estimate the total network effect of treatment (e.g., downstream alters' aggregate influence) but rather to quantify how social influence mediated by specific network layers might contribute to knowledge accuracy. Using a counterfactual prediction framework, we evaluate the reduction in accurate knowledge if a layer's influence pathways were removed. Results for primary pathways (e.g., A) are reported in the main text, while secondary pathways (e.g., A + E) are in **SI Appendix D1**. Both analyses yield consistent conclusions, though secondary pathways exhibit greater variance, possibly due to noise from misinformed intermediaries. This consistency underscores the robustness of our approach despite differing assumptions.

Third, we leverage the directionality of name generators under the assumption that influence is more likely to flow from nominated alters (B) to the nominating ego (A) than vice versa (also known as the edge-directionality test (Christakis and Fowler 2007)). This aligns with evidence that individuals often adopt behaviors or attitudes from those they explicitly recognize as peers or advisors (Valente 2012). We hypothesize that the impact of distant alters (e.g., $k$-degree separated) is mediated by critical pathways where ties follow the A→B direction (enabling A←B influence). Removing such pathways would significantly reduce knowledge accuracy, whereas pathways with reversed directed ties (A←B, where A does not nominate B, but B nominates A) would have negligible impacts. Importantly, the absence of reciprocation (B→A) suggests that the alter (B) is not similarly influenced by the ego's (A) attributes, breaking the mutual reinforcement loop that characterizes homophilous ties. Thus, by distinguishing directed and reversed directed pathways, we reduce the influence from situation where connections are driven by pre-existing similarity. See **SI Appendix F** for more discussion of modeling considerations.

Not all health knowledge interventions can generate interpersonal information transmission, particularly along extended chains of social connections. Before evaluating the relative importance of network layers in contagion, we first determine which knowledge interventions are especially effective in generating social contagion among individuals. Three interventions achieve statistical significance of social contagion at all three degrees of separation, and they are: (1) not giving baby chupones ("In your opinion when should a baby under 6 months be given a chupon?"); (2) having a birth plan ("Did you make a birth plan for this pregnancy?"); and (3) father's involvement ("In your opinion, should fathers help care for his sick children?"). Model details are available in **SI Appendix C**.

The results across three health knowledge interventions show consistent patterns in network layers' impacts on knowledge adoption. First, the results confirm the positive effects of most network layers on the adoption of correct health knowledge as their removal leads to a reduction in accuracy. For example, the social contagion of knowledge regarding giving baby chupones through the closest friend layer reveals a notable reduction in adoption rates of up to 1.16% (95% CI [1.10 %, 1.22%]). This reduction corresponds to approximately 179 individuals (=15398*1.16%) who have adopted accurate health knowledge on the practice of chupones as a result of social influence potentially transmitted via the pathways facilitated specifically by closest friends. On the contrary, the difference if closest friends were removed is 0.15% if the reversed direction of long-range chains is used in constructing critical pathways. Compared to the reduction observed with directed ties (blue dots in **Figure 5(a)**), the reductions are negligible for most layers when using reversed directionality (yellow dots). This is consistent with our hypothesis that the impact of distant alters (e.g., $k$-degree separated) is mediated by critical pathways where ties follow the A→B direction (enabling A←B influence).

[Figure 5]

Second, a pattern of consistent ordering is evident in the effects of network layers across three knowledge interventions. Network layers such as parent, patron, trusting to lend money, and discussing personal and private matters consistently exhibit either no effect or a marginal reduction in knowledge adoption when removed. In contrast, the network layer composed of closest friends exhibits significant and substantial reductions. Similarly, both the relationships involving giving and getting health advice demonstrate substantial reduction in knowledge adoption.

The scatter plot in **Figure 5(b)** reveals an approximately linear relationship between network torque (log-transformed), which is a structural property of a network layer that creates non-redundant pathways, and the resultant change in knowledge accuracy due to exogenous interventions ($\hat{\beta} = 0.83, p < 0.05$). The change, presented on the Y-axis, is averaged over knowledge accuracy reductions across the three interventions. Network layers such as "partner," and "patron," which have a limited ability to alter the underlying pathways, do not substantially influence knowledge adoption. In contrast, network layers characterized by large network torque, such as "closest friend" and "health advice give," exhibit a significant reduction in knowledge adoption if they were removed from the underlying social networks. The relationships between network torque and knowledge reduction in the experiment are also consistent when we relaxed the conditioning on primary pathways ($\hat{\beta} = 0.78, p < 0.05$) or the conditioning to include an expanded set of knowledge outcomes ($\hat{\beta} = 0.41, p < 0.005$), reported in **SI Appendix D**.

It is intriguing to note that social relationships involving giving health advice, despite displaying only moderate levels of network torque, show considerable reductions in knowledge adoption when removed, compared to relationships "trust to lend money," "free time" and "sibling," which have comparable or higher levels of network torque. Take, for example, the network torque of the relationship "free time," which falls within a similar range as the relationships involving giving health advice. Its impact on the reduction in health knowledge adoption, however, is close to zero, constituting only a small fraction of the reduction observed if the health-related relationships are removed. These findings imply that, although the multiplex network structure plays a significant role in the transmission of social influence, it only explains a limited portion of the variance. Other dimensions, such as the actual nature of the social relationship (and the content ordinarily transmitted in it, e.g., health information versus money), also play a crucial role.

**Discussion**

Here, we investigated two related objectives. We first proposed a new measure to quantify the overlap, within a broader networked population, of the network layers arising from multiplex ties, which we call "network torque." This measure captures the extent to which a specific type of relationship can create non-redundant pathways for social contagions within a defined population. Network torque focuses on how much one type of relationship contributes to non-redundant network pathways connecting distant individuals, offering direct insights into network contagion that other metrics, such as multiplex network density or network overlap, do not directly capture. This makes it especially relevant for studies examining information flow or influence propagation within bounded populations, where understanding the precise contribution of each layer to connectivity may be crucial.

We find distinct patterns in network torque across different relationship types. Moreover, three structural characteristics of social ties — overall prevalence, dyadic monoplexity, and bridgeness — are relevant to network torque. Although conceptually distinct and operating under independent logics of tie formation, these features together increase the likelihood of establishing non-redundant pathways. The relationship type with the highest network torque is "closest friend," which represents one of the most prevalent relationships in the system, accounting for 13.8% of all nominated relationships, one of the relationship types that are most likely to be monoplex ties (54.1%), and the relationship type that has highest betweenness centrality. Notably, network torque – like some other structural measures, such as modularity (Fortunato and Barthélemy 2007) and average geodesic distance (Newman 2010) – may be sensitive to network size. However, our analysis avoids this issue as we compared network torque across different network layers within the same networks rather than compared a layer's torque across different networks (**Figure 4c-d**).

Early work on network multiplexity emphasizes causal priors, e.g., changes in societal or institutional conditions (Fischer 1977), and relationship outcomes, e.g., enhancing trust, certainty and reciprocity (Gondal and McLean 2013; Powell, Koput, and Smith-Doerr 1996; Uzzi 1996), but little attention was given to how distributions of multiplex ties within a network might influence information cascades. Our work indicates that micro-level interactions, e.g., such as a respondent's choice of closest friends or with whom to share health information, can be related to macro-level patterns of social structure, influencing the overall diffusion of information and behavior within a community. And so, in our second objective, we evaluated the role of such non-redundant pathways in social contagions induced by exogenously delivered network interventions. We find that network layers with higher network torque, like "closest friend" and "health advice give," are crucial for knowledge adoption as their removal leads to a significant reduction in knowledge adoption in the village-wide networks. This is contrasted with layers like "partner" and "trust to lend money," which, due to their limited influence on pathways, show less impact on village-wide knowledge adoption.

Network-based interventions are gaining recognition as an effective approach for accelerating behavior change (Airoldi and Christakis 2024; Alexander et al. 2022; Bond et al. 2012; Centola 2010; Kim et al. 2015; Muchnik et al. 2013; Rand et al. 2011; Valente 2012), and our research contributes to the expanding body of evidence. Using an experimentally induced contagion, we found that the level of network torque appeared to align with the observed reduction in knowledge adoption if a network layer was removed. While many network targeting algorithms aim to identify structurally or socially influential individuals, with the expectation that their adoption of an innovation will lead to widespread adoption in entire communities, these approaches have their limitations. For example, identifying structurally influential individuals can be difficult, and in some cases, it requires mapping of the whole population network, which is in itself often prohibitively expensive. One exception is the friendship nomination algorithm, which leverages the friendship paradox – that is, that friends of a randomly chosen individual tend to be closer to the network's center (Kim et al. 2015; Alexander et al. 2022; Airoldi and Christakis 2024; Feld 1991). However, while the friendship nomination algorithm has proven effective in identifying influential individuals without extensive network mapping (Airoldi and Christakis 2024; Christakis and Fowler 2010; Kim et al. 2015) , the role of multiplexity, which has great impact on the global structure of a network, remains under-explored.

Our work presents a potential enhancement to network targeting, available for future work, that identifies and targets structurally conducive relationships and pathways (Valente 2012).  By focusing on the social relationships, e.g., closest friends, that are most essential for non-redundant critical pathways, it becomes more probable for information to spread far and deep within networked populations, reaching individuals located in distant network clusters. Such an approach may be especially effective in generating large-scale cascades in human social networks in rural or developing societies. These networks are typically localized and are characterized by distinct clusters of familial and friendship ties that are often segmented within the same community. Future studies can develop experimental designs that use network torque of various relationship types to generate community-wide cascades.

The importance of close friend ties for the non-redundant transmission of information is also in keeping with theories about the evolution of friendship (Tooby and Cosmides 1996; Christakis 2019). The very fact that humans have friends and the emotional apparatus to support such relationships is evidence that simple models of tit-for-tat reciprocation do not, and cannot, fully account for altruism and sociality in human groups. Friendships offer something that kin and simple exchange relationships cannot – including reliable access to novel or useful information originating further away in social networks.

**Figures and Tables**

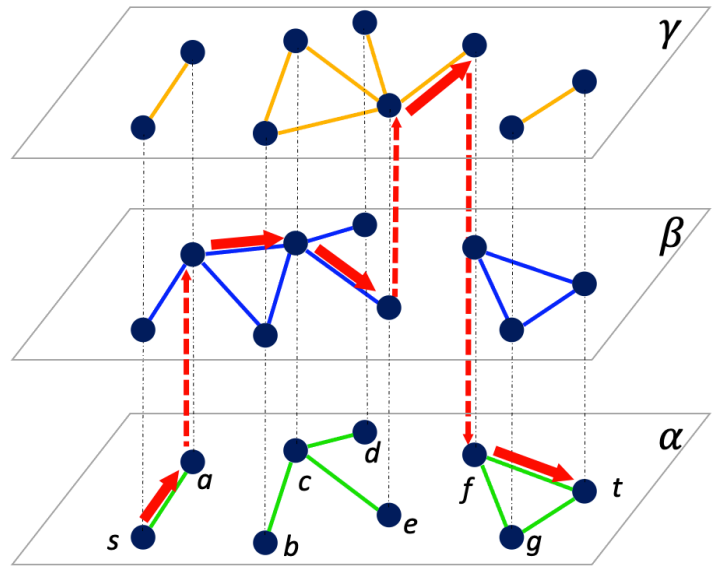

**Figure 1**. A Stylized Three-layer Pathway and Information Flow. Nodes across layers represent the same individuals, indicated by thin dashed lines, and edges with different colors represent three relationships, family (network layer $\alpha$); discussing personal and private matters (layer $\beta$); and sharing health-related information (layer $\gamma$). A new piece of information enters at node $s$, passing through a pathway traced by red arrows composed from edges from all three layers of networks, and eventually reaches the end point $t$. The 5-degree path (thick red arrows) is composed of the following sequences of edges: $s(\alpha) \rightarrow a(\alpha\text{-}\beta) \rightarrow c(\beta) \rightarrow e(\beta\text{-}\gamma) \rightarrow f(\gamma\text{-}\alpha) \rightarrow t(\alpha)$, where $e(\beta\text{-}\gamma)$ means the node $e$ shared on layers $\beta$ and $\gamma$.

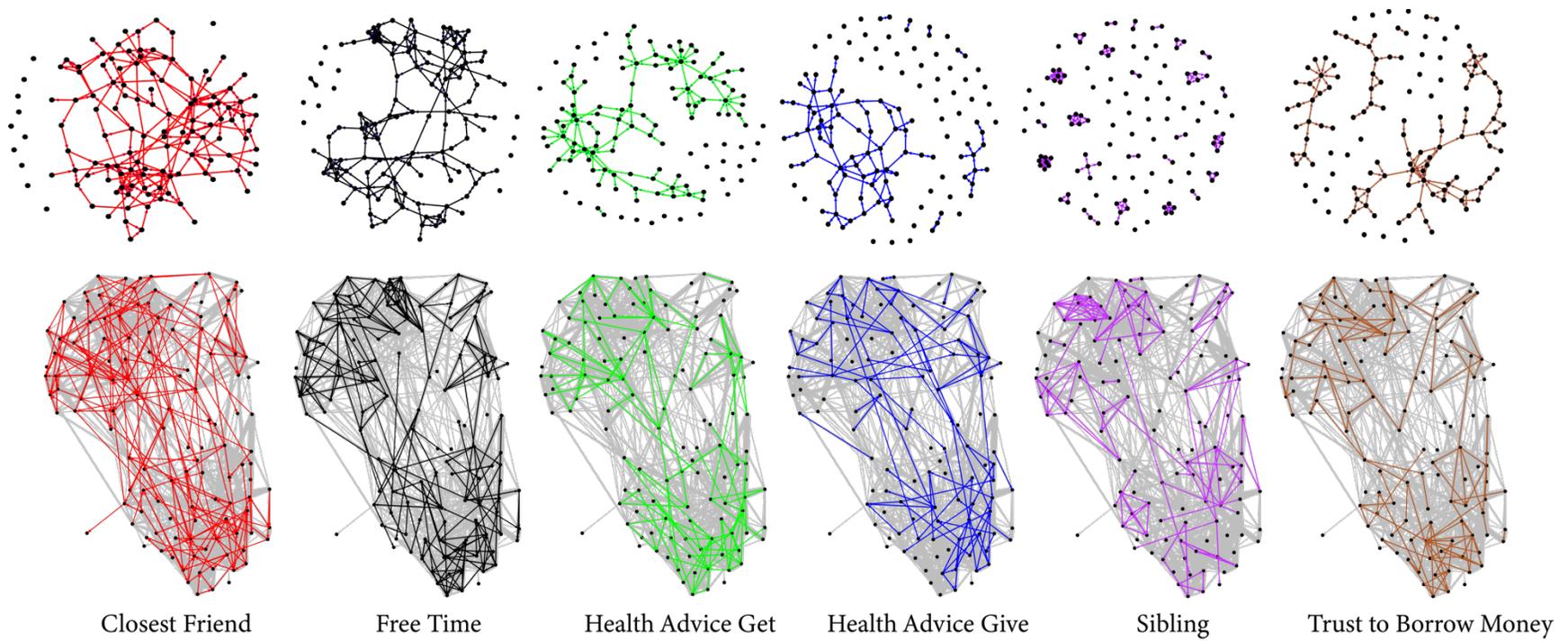

**Figure 2.** Tie Distributions and Structures of Six Layers of Network in a Single Village. The network visualizations in the upper row portray the topological structure of the network layers when examined independently. The lower row displays the structural distributions of six relationships, each denoted by a unique color, on top of the composite networks derived from all 11 types of relationships combined (gray color). The width of the ties corresponds to tie multiplexity, i.e., the number of unique relationships that exist between pairs of nodes in the network. The tie counts in network layers are as follows: 246 for the closest friend relationship, 270 for free time activities, 156 for seeking health advice, 163 for providing health advice, 197 for sibling relationships, and 152 for trust-based borrowing of money.

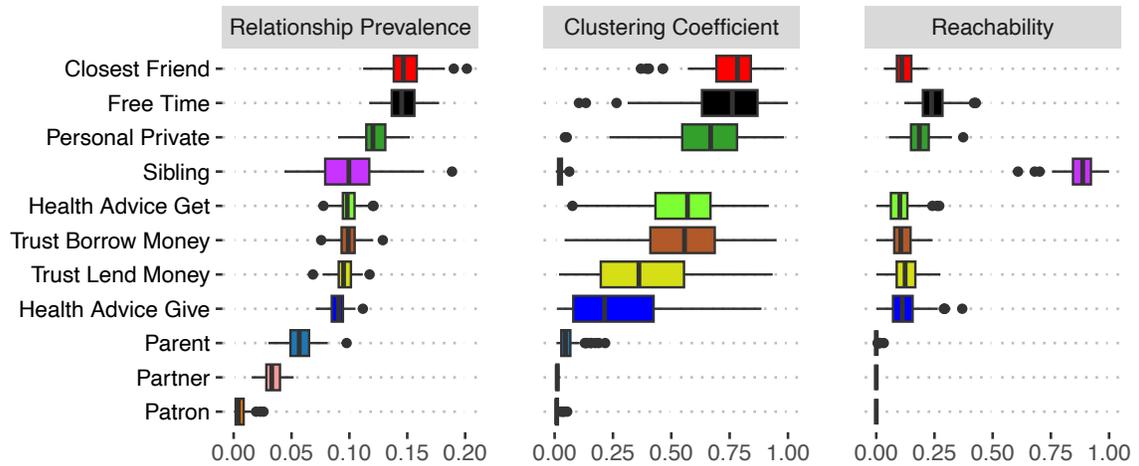

**Figure 3**. Network Descriptive Statistics of 11 Types of Relationships in 110 Villages. "Relationship prevalence" measures the proportion of ties in a village that belong to a relationship type. "Clustering coefficient" measures the degree to which nodes in a graph tend to cluster together. "Reachability" measures the likelihood that two nodes can be connected through a finite path.

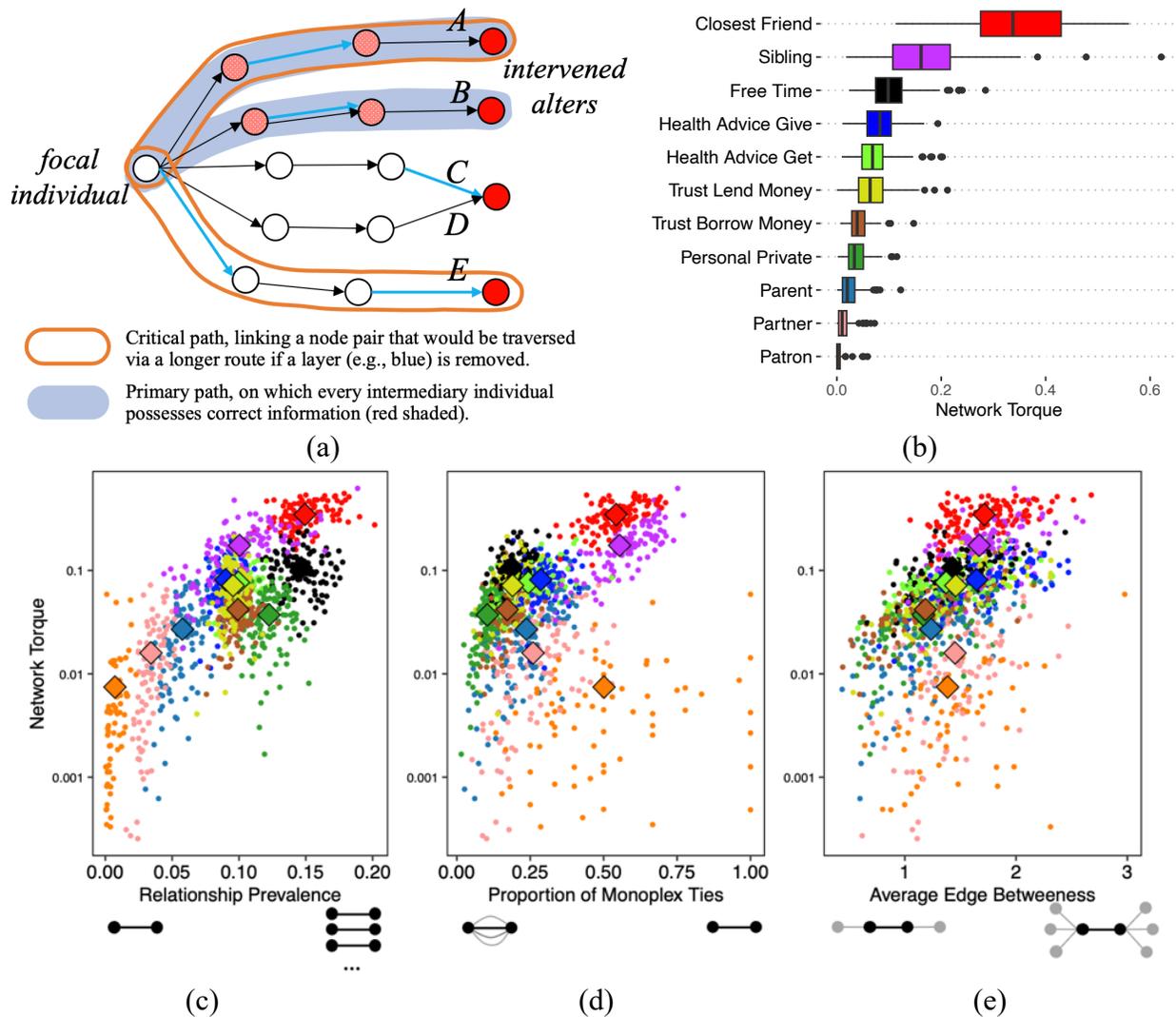

**Figure 4.** Non-redundant Network Pathways and Structural Correlates of Network Torque. (a) Possible Paths for Social Contagion. A focal individual is connected with 4 3-degree alters who have received educational modules (red shaded) at Wave 1. The intermediary nodes either possess accurate knowledge or were intervened (pink shaded) or possess incorrect knowledge (blank). Black and blue edges represent two different types of relationships. Directionality of the edge indicates nomination. *A* and *F* are critical pathways (circled in orange); *A*, *B* and *C* are primary pathways (highlighted in lightgray). (b) Box plot of the variations of network torque of 11 types of relationship (network layers) across 110 villages. (c-e) Three structural correlates of network torque: relationship prevalence, proportion of monoplex ties, and average edge betweenness centrality. Circles represent village network layers, stratified by color based on their respective layers; each diamond corresponds to the averaged values of a specific network layer across 110 villages. Relationship types are color coded consistently across the panels.

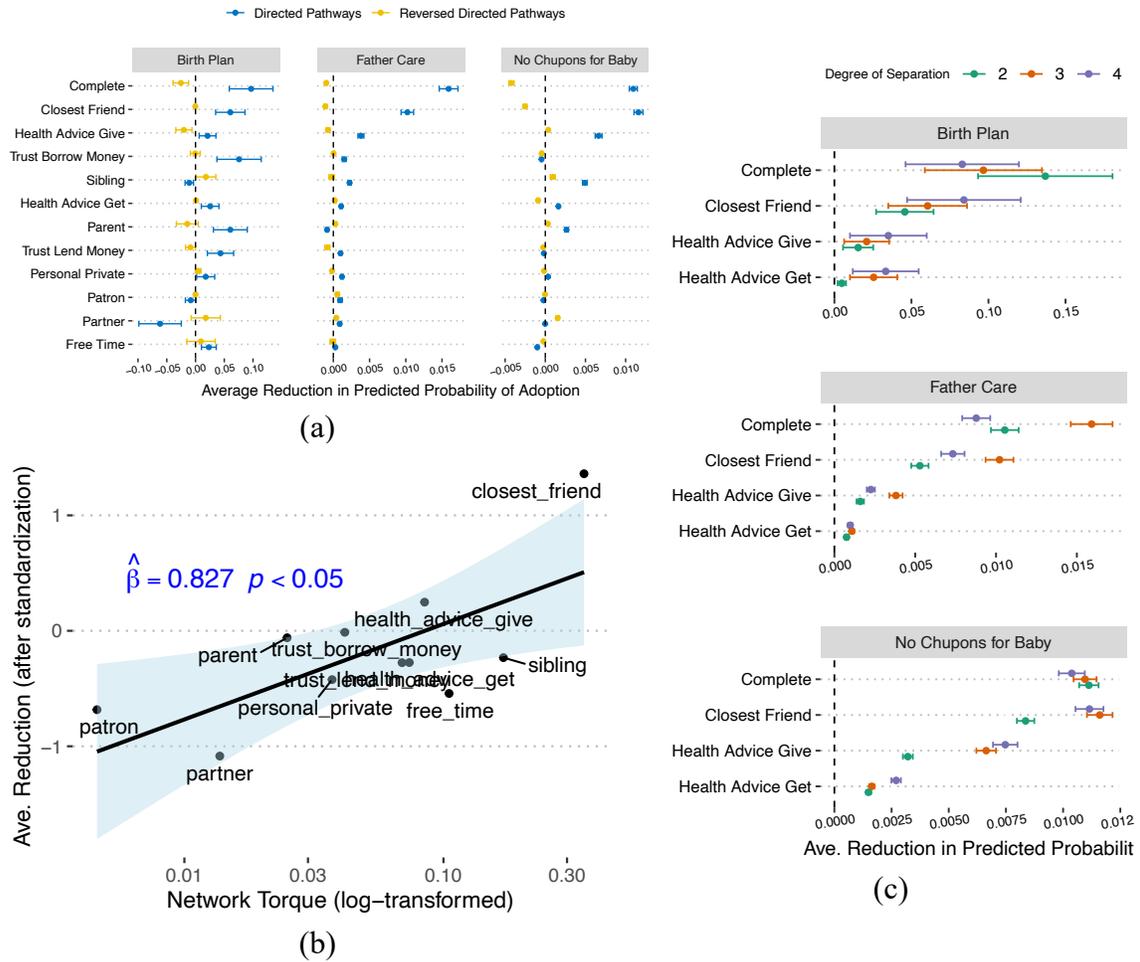

**Figure 5.** Change in Knowledge Accuracy Due to Exogenous Interventions. (a) Average reduction in predicted probability of accurate health knowledge due to 3-degree social influence on respective network layers. The bars on the point estimates are 99% confidence intervals. The X-axis reports the difference in predicted outcomes when a network layer is removed versus when it is included in the model. Here, only those influences passing through 3-degree primary pathways are reported; the SI Appendix includes results of other conditions. (b) Reduction in predicted probability and network torque. The Y-axis represents the average reduction (after standardization) if a network layer is removed, across three health interventions. The X-axis quantifies log-transformed network torque. The lightgray band indicates 95% confidence interval. (c) Average reduction in predicted probability of accurate health knowledge through selected network layers by degree of separation.

**Table 1**. The Name Generator Battery

| Name Generator | Short Name | Category |
|---|---|---|
| **What is the name of your mother? What is the name of your father?** | Parent | Role relation |
| **What are the names of your siblings over the age of 12 that live or work here?** | Sibling | Role relation |
| **Are you married or living in a civil union? What is the name of your partner?** | Partner | Role relation |
| **Do you have a patrón/patrona?** | Patron | Role relation |
| **Who do you trust to talk to about something personal or private?** | Personal and Private | Friendship |
| **With whom do you spend free time?** | Free Time | Friendship |
| **Besides your partner, parents or siblings, who do you consider to be your closest friends?** | Closest Friend | Friendship |
| **Who would you feel comfortable asking to borrow 200 lempiras from if you needed them for the day?** | Trust to Borrow Money | Exchange |
| **Who do you think would be comfortable asking you to borrow 200 lempiras for the day?** | Trust to Lend Money | Exchange |
| **Who would you ask for advice about health-related matters?** | Health Advice Give | Exchange |
| **Who comes to you for health advice?** | Health Advice Get | Exchange |